\documentclass[aps,pra,twocolumn,amsmath,amssymb,,showpacs,superscriptaddress]{revtex4-1}
\usepackage{graphicx}% Include figure files
\usepackage{dcolumn}% Align table columns on decimal point
\usepackage[mathlines]{lineno}
\usepackage{hyperref}
\usepackage{bm}% bold math
\usepackage{epstopdf}
\usepackage{epsfig}

\begin{document}

\title{Two-photon optics of Bessel-Gaussian modes}

%\author{Melanie~McLaren,$^{1,2}$ Jacquiline~Romero,$^{3}$ Miles~J.~Padgett,$^{3}$ Filippus~S.~Roux,$^1$ and Andrew~Forbes$^{1,2*}$}
\author{Melanie~McLaren}
\affiliation{CSIR National Laser Centre, P.O. Box 395, Pretoria 0001, South Africa} 
\affiliation{Laser Research Institute, University of Stellenbosch, Stellenbosch 7602, South Africa}
\author{Jacquiline~Romero}
\affiliation {Department of Physics and Astronomy, SUPA, University of Glasgow, Glasgow, UK}
\author{Miles~J.~Padgett}
\affiliation {Department of Physics and Astronomy, SUPA, University of Glasgow, Glasgow, UK}
\author{Filippus~S.~Roux}
\affiliation{CSIR National Laser Centre, P.O. Box 395, Pretoria 0001, South Africa} 
\author{Andrew~Forbes}
\email[Corresponding author: ]{aforbes1@csir.co.za}
\affiliation{CSIR National Laser Centre, P.O. Box 395, Pretoria 0001, South Africa} 
\affiliation{Laser Research Institute, University of Stellenbosch, Stellenbosch 7602, South Africa}

\date{\today}

\begin{abstract}
In this paper we consider geometrical two-photon optics of Bessel-Gaussian modes generated in spontaneous parameteric down-conversion of a Gaussian pump beam.  We provide a general theoretical expression for the orbital angular momentum (OAM) spectrum and Schmidt number in this basis and show how this may be varied by control over the radial degree of freedom, a continuous parameter in Bessel-Gaussian modes.  As a test we first implement a back-projection technique to classically predict, by experiment, the quantum correlations for Bessel-Gaussian modes produced by three holographic masks, a blazed axicon, binary axicon and a binary Bessel function. We then proceed to test the theory on the down-converted photons using the binary Bessel mask.  We experimentally quantify the number of usable OAM modes and confirm the theoretical prediction of a flattening in the OAM spectrum and a concomitant increase in the OAM bandwidth.  The results have implications for the control of dimensionality in quantum states. 
\end{abstract}

% insert suggested PACS numbers in braces on next line
% insert suggested keywords - APS authors don't need to do this
%\keywords{}

%\maketitle must follow title, authors, abstract, \pacs, and \keywords
\maketitle

\section{Introduction}
Quantum entanglement has formed the basis of several quantum information technologies, including quantum computing and quantum communication. Such examples include quantum ghost imaging \cite{Pittman-1995}, quantum cryptography \cite{Ekert1991, Gisin2002}, and quantum computing algorithms \cite{Nielsen2000}. The amount of information in an entangled state depends on the dimension of its associated Hilbert space. Photonic quantum information is often encoded in photon polarisation, which is constrained in a two-dimensional Hilbert space. In contrast, the spatial degrees of freedom (transverse spatial modal profile) of a photon has an infinite dimensional Hilbert space. Researchers are therefore focussing on the spatial modes of paraxial optical beams to increase the information capacity per photon. To this end, the most commonly used basis is that of the Laguerre-Gaussian (LG) modes.
\\
It was shown that the LG modes are orbital angular momentum (OAM) eigenstates of photons \cite{Allen1992}. Each photon in such a beam carries an amount of OAM equal to $\ell \hbar$, where $\ell$ is the azimuthal index of the LG mode. OAM is conserved in spontaneous parametric down-conversion (SPDC); this has been demonstrated both theoretically \cite{Arnaut2000, Franke2002} and experimentally \cite{Mair2001}. The implication thereof is that a pair of down-converted photons are naturally entangled in terms of the OAM eigenstates. Although LG modes are also characterised by a radial index \textit{p} this is often set to zero as higher radial indices require complex amplitude modulation (intensity masking) \cite{Arrizon-2003}, which results in the loss of optical power.
\\
By approximating the phase matching condition with a Gaussian function, it was shown \cite{Law2004} that the Schmidt basis for the quantum state produced in SPDC is the LG modes. Without this approximation, the LG modes are still close to being the Schmidt basis \cite{Salakhutdinov2012}. Thus, the LG modes are not only entangled in the azimuthal index, but also in terms of the radial index. As the azimuthal component of LG modes is orthogonal in $\ell$ and OAM is conserved in SPDC, the exact Schmidt basis for the down-converted quantum state would be a basis of OAM eigenstates. High-dimensional entanglement is dependent on the number of usable OAM modes in the state \cite{Torres2003}. However, the experimental parameters (e.g. mode size of single-mode fiber) involved in the detection of the spectrum of LG modes place restrictions on the control one has over the bandwidth of OAM components in the entangled state. 
\\
The LG modes are not the only OAM basis. Higher-order Bessel beams \cite{Durnin1987, Durnin21987} and Bessel-Gaussian (BG) beams \cite{Gori1987} also have helical wavefronts and carry OAM. The spatial modal bases are related by unitary transformations such that the down-converted quantum states are also entangled in terms of the BG modes. The BG modes allow for additional control over quantum state preparation as they have a continuous radial scale parameter that distinguishes different modes, instead of a discrete radial index as in LG modes.
\\
OAM entanglement in the BG basis has already been successfully demonstrated \cite{McLaren2012}. In this paper we examine the methods in which OAM entanglement may be measured in the BG basis. We introduce back-projection as a tool to study BG projective measurements, using two-photon geometric optics to predict the strength of the coincidence correlations. We also quantify the number of measurable OAM modes by calculating the Schmidt number, and demonstrate a clear dependence on the radial component. This suggests a means to increase the dimensionality of entangled states.

\section{Theory}
The probability for the biphoton quantum state after the SPDC process to contain a particular measurement state $\rho_{\textrm{m}} = \left|\Psi_{s}\right\rangle \left|\Psi_{i}\right\rangle \left\langle\Psi_{s} \right| \left\langle \Psi_{i} \right|$, is given by the trace $\textrm{Tr} \{ \rho \rho_{\textrm{m}} \} = \left| {\cal M} \right|^{2}$, where ${\cal M}$ is the scattering amplitude. For monochromatic paraxial pump, signal and idler beams in a degenerate collinear SPDC process with type I phase matching, the scattering amplitude is given by \cite{Miatto2011}
\begin{eqnarray}
{\cal M} & = & \Omega_0 \int M_{s}^*({\textbf{K}}_1) M_{ i}^*({\textbf{K}}_2) M_{ p}({\textbf{K}}_1 + {\textbf{K}}_2) \nonumber \\ & & \times P \left({\Delta k_z}\right)\ {\textrm{d}^2 k_1\over (2\pi)^2}\ {\textrm{d}^2 k_2\over (2\pi)^2},
\label{mamp2}
\end{eqnarray}
where $M_{s} (\textbf{K})$, $M_{i} (\textbf{K})$ and $M_{p} (\textbf{K})$ are the Fourier spectra of the two-dimensional mode profiles for the signal, idler and pump beams, respectively; $\textbf{K}$ represents the coordinate vector in the two-dimensional transverse Fourier domain; $\Omega_0$ is a constant that determines the overall conversion efficiency; and $P \left({\Delta k_z}\right)$ is a function that represents the phase matching condition.
\\
The phase matching condition is given in terms of a sinc-function
\begin{equation}
P\left({\Delta k_z}\right) = \textrm{sinc}\left(\frac{\Delta k_z L}{2 \pi}\right) = \textrm{sinc}\left(\zeta \left|\textbf{K}_{1} - \textbf{K}_{2}\right|^{2}\right),
\label{sinc}
\end{equation}
where 
\begin{equation}
\zeta = \frac{n_{o}\lambda_{p}L}{8\pi^{2}},
\label{zeta}
\end{equation}
with $n_{o}$ being the ordinary refractive index of the non-linear crystal, $\lambda_{p}$ being the wavelength of the pump and $L$ being the crystal length. Assuming the pump wavelength is much smaller than any of the other dimension parameters, the width of the sinc-function in Eq.~(\ref{zeta}), as determined by $\zeta^{-1/2}$, is much larger than the widths of the angular spectra of the pump, signal or idler beams \cite{Torres2003}. Hence, one can approximate $P=1$ and the Fourier integral in Eq.~(\ref{mamp2}) can be transformed into a spatial domain given by
\begin{eqnarray}
{\cal M} & = & \Omega_0 \int m_{s}^*({\textbf{x}}) m_{ i}^*({\textbf{x}}) m_{ p}({\textbf{x}}) {\textrm{d}^2 x},
\label{eq:spatial}
\end{eqnarray}
where $m_{s}(\textbf{x})$, $m_{i}(\textbf{x})$ and $m_{p}(\textbf{x})$ are the two-dimensional mode profile functions for the signal, idler and pump beams, respectively.
\begin{figure}
\centering
\includegraphics[scale=0.38]{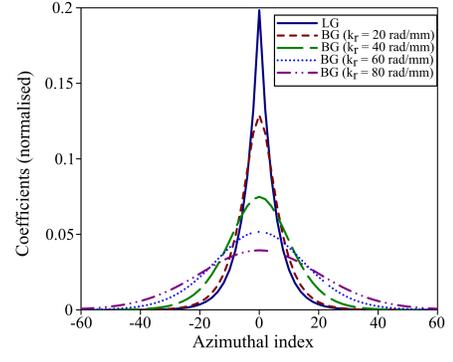}
\caption{Comparison of the OAM spectra for the LG modes ($k_{r}=0$) and BG modes ($k_{r} = 20, 40, 60, 80$ rad/mm) using Eqs.~(\ref{LGcoeff}) and~(\ref{BGcoeff}), respectively, with $\omega_{0} = 0.5$ mm and $\omega_{1} = 0.23$ mm. The number of usable OAM modes increases with the radial wavevector $k_{r}$.}
\label{fig:besselbw}
\end{figure}
The pump beam has a mode profile described by a Gaussian function, which is expressed as
\begin{equation}
m_{p} = \frac{1}{\omega_{0}} \sqrt{\frac{2}{\pi}} \exp\left[ -\frac{(x^{2} +y^{2})}{\omega_{0}^{2}}\right] ,
\label{Gauss}
\end{equation}
where the radius of the mode profile of the pump beam is given by $\omega_{0}$. We consider the case where the signal and idler beams are BG modes with azimuthal indices $\ell$ and $-\ell$, respectively, and with scaling parameters $k_{r1}$ and $k_{r2}$, respectively. For simplicity we assume that $k_{r1} = k_{r2} = k_{r}$. 
A BG mode with a specific $\ell$-value is produced by evaluating the following integral,
\begin{equation}
M^{\textrm{BG}}_{\ell} = {1\over 2\pi} \int_0^{2\pi} {\cal G} \exp(-i\ell\beta)\ \textrm{d}\beta .
\label{gbgx}
\end{equation}
The generating function, ${\cal G}$ for BG modes at $z = 0$ is then given by
\begin{eqnarray}
{\cal G} & = & \sqrt{\frac{2}{\pi}} \frac{1}{\omega_{1}}  \exp \{ik_{r} [y \cos(\beta) -x \sin(\beta)]\} \nonumber \\ & & \times \exp \left[  - \frac{(x^{2} + y^{2})}{\omega_{1}^{2}} \right],
\label{Gen}
\end{eqnarray}
where the radius of the Gaussian envelope of the mode is $\omega_{1}$ and $\beta$ is an angular generating parameter. 
\\
The coefficients for any given value of the azimuthal index $\ell$ (with opposite signs for the signal and idler beams, respectively) can be extracted by substituting Eq.~(\ref{Gauss}) and Eq.~(\ref{Gen}) into Eq.~(\ref{eq:spatial}), and solving the integral. The OAM spectrum, represented by these coefficients, is given by
\begin{eqnarray}
C_{\ell}& = & (-1)^{\ell} \sqrt{\frac{2}{\pi}}\frac{2\Omega_{0}\omega_{0}^{2}}{2\omega_{0}^{2} + \omega_{1}^{2}} \nonumber\\&&\times \exp\left[\frac{-k_{r}^{2}\omega_{1}^{4}}{4(2\omega_{0}^{2}+\omega_{1}^{2})}\right] \frac{I_{\ell}\left[\frac{k_{r}^{2}\omega_{0}^{2}\omega_{1}^{2}}{2(2\omega_{0}^{2}+\omega_{1}^{2})}\right]}{I_{\ell}\left[\frac{k_{r}^{2}\omega_{1}^{2}}{4}\right]},
\label{BGcoeff}
\end{eqnarray}
where $I_{\ell}(\cdot)$ is the modified Bessel function of the first kind \cite{Abramowitz1972}. The equivalent coefficients in the LG basis, for zero radial index, are given by
\begin{equation}
C_{\ell} = \Omega_{0} \sqrt{\frac{2}{\pi}} \left( \frac{2 \omega_{0}^{2}}{2\omega_{0}^{2} + \omega_{1}^{2}} \right)^{\left|\ell\right| + 1}.
\label{LGcoeff}
\end{equation}
An estimate of the OAM bandwidth of this spectrum can be calculated by computing the Schmidt number \cite{Law2004} given by
\begin{equation}
K = \frac{\left(\sum\limits_{\ell}C_{\ell}^{2}\right)^{2} }{\sum\limits_{\ell}C_{\ell}^{4}}.
\label{schmidt}
\end{equation}
In the case where the OAM spectrum is computed in terms of the BG modes, one cannot obtain a closed form expression for the Schmidt number. However, one can compute the Schmidt number numerically from the analytical result in Eq.~(\ref{BGcoeff}) for any given value of $k_{r}$. Figure.~\ref{fig:besselbw} shows the OAM spetra for the LG and BG modes. In the case where Eq.~(\ref{sinc}) is not approximated as 1, the OAM spectra are limited by the length of the non-linear crystal \cite{Miatto2012}.

\section{Simulation of entanglement with classical light}
There has been a great deal of interest in mathematically determining a method to predict quantum correlations in particular reference to quantum communication and imaging \cite{Barnett2000a, Barnett2000b, Tan2003}. The measurable correlation of the entangled photons in a typical SPDC experiment depends on the quality of state generation (e.g. the range of OAM states that SPDC actually produces) and the state detection (e.g. the range of the OAM states that can be detected by the measurement scheme).  It is useful to isolate the effect of generation from detection, and vice versa. We are interested in investigating the quality of our detection system, and so to this end, we introduce a back-projection experiment inspired by the advanced-wave representation of Klyshko \cite{Klyshko1988}.
\\
The Klyshko picture is useful in assessing the conditional probability distribution--  the probability of detecting a photon at detector B given that another photon is detected at detector A. Klyshko considered the field detected in arm A as propagating in reverse back to the crystal plane where it reflects off the crystal to propagate forward through the system to detector B. Using this picture and geometrical optics arguments, the two-photon correlations measured in SPDC can be predicted, as in the ghost imaging and two-photon optics experiments in \cite{Pittman-1995, Pittman1996}. 
\begin{figure}[htbp]
\centering
\includegraphics[scale=0.35]{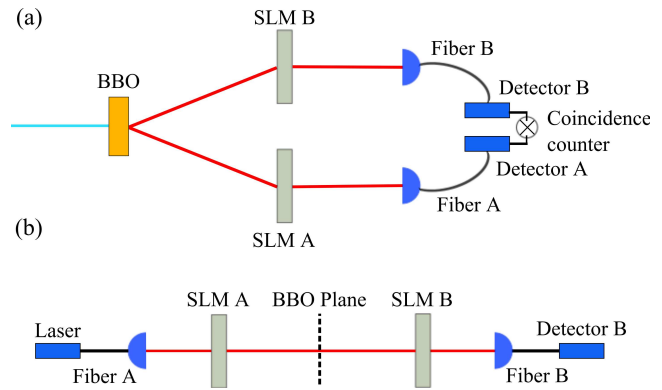}
\caption{(a) Schematic of an entanglement setup. The entangled photon pairs were generated by the BBO crystal, the combination of the SLMs and detectors projected the photon pair into a particular state and the detection of the pairs was measured with a coincidence counter. (b) Unfolded diagram of (a), where the BBO crystal is replaced with a mirror such that SLM A is imaged onto SLM B. Light from a 710 nm diode laser was coupled into fiber A, where after it was imaged to SLM A followed by SLM B and then re-coupled into fiber B. By placing a CCD camera at the plane of the BBO crystal, we can guarantee that we are measuring BG fields.}
\label{fig:BPscheme}
\end{figure}
More than a theoretical tool, the Klyshko picture can also be applied experimentally.  One of the detectors can be replaced with a classical light source and propagated through one arm back onto the crystal plane, where a mirror has been placed (this corresponds to the wave propagating in reverse).  At the crystal, this back-projected beam was reflected and propagated forward onto the components of the other arm and onto the other detector.  The number of photons registered by this detector can be optimised to ensure the stringent alignment required by the system, and more importantly, can be used to predict the expected behaviour of the two-photon correlation. With these in mind, we implemented a back-projection experiment. Figure~\ref{fig:BPscheme}(a) shows a simple schematic of an entanglement setup.

An unfolded setup of Fig.~\ref{fig:BPscheme}(a) is shown in Fig.~\ref{fig:BPscheme}(b). A 710 nm diode laser with a Gaussian profile replaced detector A and was connected to fiber A.  The output was imaged through the system to SLM A, which was then imaged onto a mirror at plane of the crystal.  From here, the light was imaged onto SLM B, and SLM B is in turn imaged onto the facet of single-mode fiber B. The fiber was coupled to detector B which registered the single photon count rate.  In order to have significant single photon counts, a careful choice of each phase pattern must be made.  These patterns should ensure the fundamental mode from fiber A is coupled into fiber B. To illustrate this, if a positive lens function is encoded onto SLM A,  a negative lens function must be encoded onto SLM B to produce a Gaussian mode which can only then be coupled into fiber B. This can be seen mathematically using ABCD matrices \cite{Belanger91}:
\begin{eqnarray}
\left[ \begin{array}{c} x_2 \\ \alpha_2 \end{array} \right] & = & \begin{bmatrix} 1 & 0 \\ -1/f_{2} & 1 \end{bmatrix} \begin{bmatrix} -1 & 0 \\ 1/f_{1} & -1 \end{bmatrix} \left[ \begin{array}{c} x_1 \\ \alpha_1 \end{array} \right] \nonumber\\&&
 =  \begin{bmatrix} -1 & 0 \\ \left(1/f_{2}+1/f_{1}\right) & -1 \end{bmatrix}  \left[ \begin{array}{c} x_1 \\ \alpha_1 \end{array} \right]
\label{ABCD}
\end{eqnarray}
When $f_{1} = -f_{2}$ the transverse and angular positions of the initial and final beam remain the same. However, identical focal lengths result in a change in the angular position, producing a divergent beam at fiber B and thus reducing the coupling efficiency.  In the context of the spatial modes, which we are trying to measure, maximum coupling of the light from fiber A to fiber B occurs when the transmission functions encoded on the SLMs are phase-conjugates of each other.

\section{Back-projection results}
OAM entanglement is typically measured in the LG basis by encoding only the azimuthal phase term onto an SLM, described by the transmission function:
\begin{equation}
T(\phi) =  \exp(i\ell \phi),
\label{LG}
\end{equation}
\noindent where $\phi$ is the azimuthal angle. Depending on the azimuthal index $\ell$ an incoming Gaussian mode is transformed into an approximated LG mode carrying OAM of $\ell \hbar$ per photon. This is only an approximation as the radial components of the LG function have been neglected in favour of efficiency \cite{Salakhutdinov2012}. As this process is reversible, a mode generated in SPDC may be converted into a Gaussian mode using the same transmission function in Eq.~(\ref{LG}). Due to the omitted radial modes, some light is lost from scattering into higher-order modes \cite{Miatto2011}. By selecting the BG basis in which to measure OAM entanglement, we have access to a continuous scaling parameter for the radial component of the BG modes. We consider blazed axicons, which have been well documented for producing Bessel-Gauss beams, binary axicons and binary Bessel functions.

\subsection{Blazed axicon}
\begin{figure}[htbp]
\centering
\includegraphics[scale=0.29]{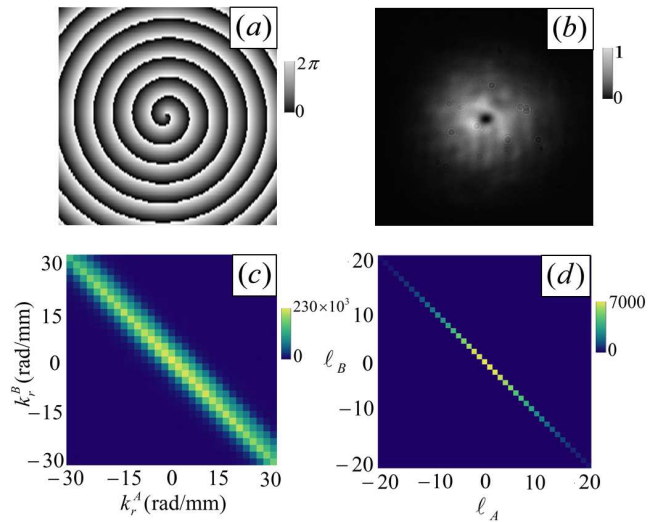}
\caption{(a) Phase pattern used to define an axicon of $k_{r} = 21$ rad/mm and $\ell = 1$. (b) CCD image of a BG beam generated from a blazed axicon function of $k_{r} = 21$ rad/mm and $\ell = 1$ at the plane of the crystal. (c) Density plot of the single count rates measured in back-projection for different blazed axicon phase patterns; varying $k_{r}$ with $\ell = 0$. (d) Density plot of the single count rates measured in back-projection for a particular blazed axicon; $k_{r} = 21$ rad/mm and varying $\ell$.}
\label{fig:holoblaz}
\end{figure}

\begin{figure}[htbp]
\centering
\includegraphics[scale=0.37]{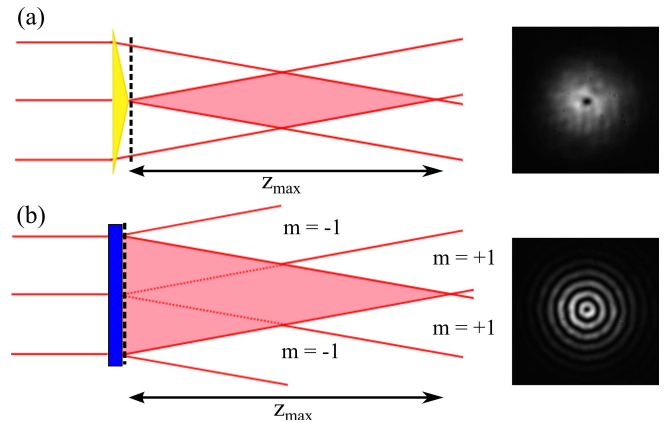}
\caption{(a) Formation of a Bessel-Gaussian beam using a standard blazed axicon. The CCD image shows the plane of initial formation (illustrated by the dashed black line), where no rings are seen. (b) Formation of a Bessel-Gaussian mode using a binary axicon. The CCD image shows the initial plane (illustrated by the dashed black line) where a Bessel-like beam is observed. However, it is not a true BG beam as the interference is only between the diffraction orders of the binary hologram.}
\label{fig:BessDia}
\end{figure}

The first phase pattern investigated was that of an axicon described by a blazed (kinoform element) function, first described by \citet{Turunen1988, Cottrell2007}. The conversion from Gaussian to BG mode was performed using the phase-only hologram described by the transmission function
\begin{equation}
T_{1}(k_{r}, \phi) = \exp(ik_{r}r) \exp(i\ell \phi),
\label{blazed}
\end{equation}
where $k_{r}$ is the radial wavevector and $\ell$ is the azimuthal index. The number of rings of the BG beam increases with $k_{r}$. This kinoform diffracts approximately 100\% of the incoming light into the first order. 
Figure~\ref{fig:holoblaz}(a) shows an example of such a phase pattern for $k_{r} = 21$ rad/mm and $\ell = 1$.
By placing a CCD camera in the plane of the crystal, we were able to view the mode at this plane. The shape of the beam imaged from the blazed axicon for $k_{r} = 21$ rad/mm and $\ell = 1$ is shown in Fig.~\ref{fig:holoblaz}(b). The image does not exhibit a well-defined Bessel beam. A distinct BG beam will typically form after propagating some distance after an axicon, while Fig.~\ref{fig:holoblaz}(b) only shows the beginning of the BG beam, before propagation. However, a spot of zero intensity can be clearly seen, indicative of an azimuthal phase term of non-zero $\ell$. This is more clearly illustrated in Fig.~\ref{fig:BessDia}. 
\\
Figure~\ref{fig:holoblaz}(c) shows the experimental measurements of the single counts measured at detector B. The counts recorded show a strong correlation along the diagonal corresponding to values of $k_{r}$ of equal magnitude but opposite sign. The single counts level was significant only when the system consisted of a positive radial wavevector $k_{r}$ on one SLM with the corresponding negative radial wavevector $-k_{r}$ on the other (although the radial wavevector is a positive entity, we assign a negative value to $k_{r}$ to represent the conjugate phase). This translates to a positive axicon imaged onto a negative axicon, to produce a Gaussian beam, which is analogous to the lens functions in Eq.~(\ref{ABCD}). As $k_{r}$ is a continuous variable, we expected a gradual decrease in the count rate moving away from the diagonal elements. The OAM correlations for a particular blazed axicon function are shown in Fig.~\ref{fig:holoblaz}(d).

\subsection{Binary axicon}
\begin{figure}[htbp]
\centering
\includegraphics[scale=0.29]{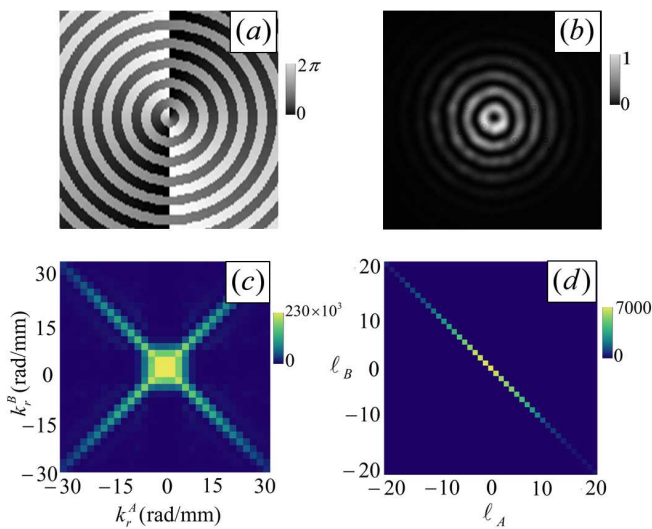}
\caption{\label{fig:holobin}(a) Phase pattern used to define a binary axicon of $k_{r} = 21$ rad/mm and $\ell = 1$. (b) CCD image of a BG beam generated from a binary axicon function of $k_{r} = 21$ rad/mm and $\ell = 1$ at the plane of the crystal. (c) Density plot of the single count rates measured in back-projection for different binary axicon phase patterns; varying $k_{r}$ with $\ell = 0$. The photons from SLM A assume a value of either positive or negative $k_{r}$ from the binary function. SLM B was encoded with the same function, which allowed BG modes of either positive or negative radial wavevectors to be converted to a Gaussian mode. (d) Density plot of the single count rates measured in back-projection for a particular binary axicon; $k_{r} = 21$ rad/mm and varying $\ell$.}
\end{figure}

One can question whether the OAM modes are truly measured in the BG basis with a blazed axicon, as the image at the crystal plane did not resemble a Bessel beam. This issue can be remedied by using a different approximation to an axicon function. That is, the second phase pattern studied also incorporated the axicon function, but as a binary function:
\begin{equation}
T_{2}(r, \phi) = {\rm sign}\left\{ \exp(ik_{r}r)\right\} \exp(i\ell\phi) ,
\label{binary}
\end{equation}
where sign$\{\cdot\}$ denotes the sign-function. The kinoform initially used to approximate an axicon was replaced with a two-level binary approximation. The efficiency of a kinoform DOE into the first diffraction order is almost $100\%$, while the efficiency of a binary function is about half that; $42\%$ in both the $m=\pm 1$ orders. The binary hologram deflects both diffraction orders symmetrically such that they interfere with each other and produce a Bessel-like region immediately after the SLM in Fig.~\ref{fig:BessDia}(b). By filtering the higher diffraction orders, a clear image of a BG mode was recorded at the crystal plane, see Fig.~\ref{fig:holobin}(b). 
\\
Similar to the previous case, an incoming Gaussian beam can be converted into a mode with a radial wavevector of either $k_{r}$ or $-k_{r}$. The recorded single count rate shown in Fig.~\ref{fig:holobin}(c) illustrates a distinct difference between the blazed and binary axicon functions. The binary function on SLM A transforms the incoming Gaussian mode into a BG mode with radial wavevector of either $k_{r}$ or $-k_{r}$, such that there is an equal probability of generating a photon with a positive or negative $k_{r}$ value. Subsequently SLM B, encoded with the same function, is also able to convert both radial vector modes into a Gaussian mode. As a result, single count rates were observed for $k_{r}^{A} = k_{r}^{B}$ and $k_{r}^{A} = -k_{r}^{B}$. The OAM correlations for a particular binary axicon function are shown in Fig.~\ref{fig:holobin}(d).

\subsection{Binary Bessel function}
\begin{figure}[htbp]
\centering
\includegraphics[scale=0.29]{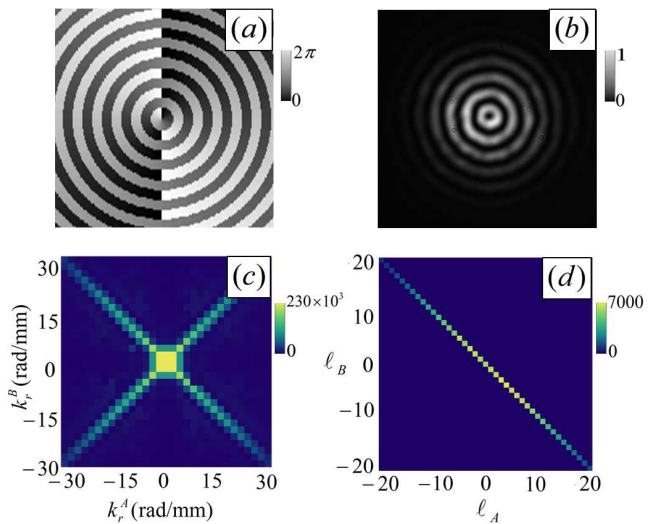}
\caption{(a) Phase pattern used to define a binary Bessel function of $k_{r} = 21$ rad/mm and $\ell = 1$. (b) CCD image of a BG beam generated from a binary Bessel function of $k_{r} = 21$ rad/mm and $\ell = 1$ at the plane of the crystal. (c) Density plot of the single count rates measured in back-projection for different binary Bessel phase patterns; varying $k_{r}$ with $\ell = 0$. (d) Density plot of the single count rates measured in back-projection for a particular binary Bessel function; $k_{r} = 21$ rad/mm and varying $\ell$.}
\label{fig:holobes}
\end{figure}

The implementation of the binary axicon function confirmed that we indeed measured BG modes. However, the axicon function acts only as an approximation to the Bessel function. Therefore, the final phase pattern considered was that of a binary Bessel function:
\begin{equation}
T_{3}(r, \phi) = {\rm sign}\left\{ {\rm J}_{\ell}(k_{r}r)\right\} \exp(i\ell\phi).
\label{bessel}
\end{equation}
Here ${\rm J}_{\ell}(\cdot)$ is the Bessel function of the first kind. Although very similar to the binary axicon function, Eq.~(\ref{bessel}) provides a more accurate description of an ideal Bessel beam. The spacing between the rings of a Bessel beam generated from an axicon remain constant with radial position, while the spaces of a theoretical Bessel beam vary in size as we move radially outward from the centre, and subsequently depend on $k_{r}$. The phase pattern and CCD image of a binary Bessel function are shown in Fig.~\ref{fig:holobes}.

A measurable count rate is again obtained along the diagonal where $k_{r}^{A} = \pm k_{r}^{B}$. However, the off-diagonal crosstalk is now less prominent, when compared with Fig.~\ref{fig:holobin}(c),  particularly surrounding $k_{r}^{A,B} = 0$. We have generated a BG mode with a better radial approximation, thus creating less overlap between the different radial wavevectors. The OAM correlations for a particular binary Bessel function are shown in Fig.~\ref{fig:holobes}(d).

\section{Down-conversion experiment and results}
In spontaneous parametric down-conversion (SPDC), a nonlinear crystal pumped with a laser beam generates pairs of entangled photons, which are then separately detected. Because we are interested in measuring spatial transverse profiles, we use a spatial light modulator (SLM) which allows arbitrary phase transformations to be performed on an incident beam. The transmission functions required in our experiment are conveniently programmed into SLMs. Moreover, the SLMs work in both the single-photon (as in SPDC) and classical (as in back-projection) regimes.
\begin{figure}[htbp]
\centering
\includegraphics[scale=0.275]{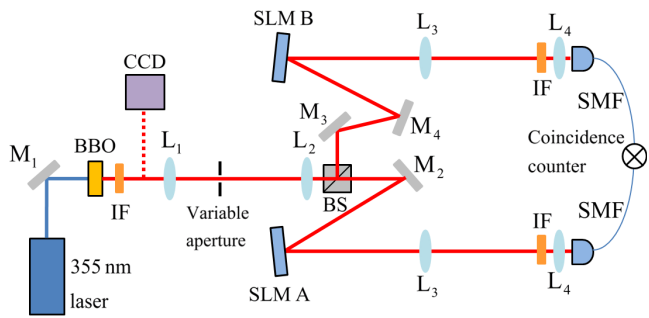}
\caption{Experimental setup used to detect the OAM eigenstate after SPDC. The plane of the crystal was relay imaged onto two separate SLMs using lenses, $\textrm{L}_{1}$ and $\textrm{L}_{2}$ ($\textrm{f}_{1} = 200$ mm and $\textrm{f}_{2} = 400$ mm), where the BG modes were selected. Lenses $\textrm{L}_{3}$ and $\textrm{L}_{4}$ ($ \textrm{f}_{3} = 500$ mm and $\textrm{f}_{4} = 2$ mm) were used to relay image the SLM planes through 10 nm bandwidth interference filters (IF) to the inputs of the single-mode fibers (SMF).}
\label{fig:setup}
\end{figure}

Our SPDC setup is shown in Fig.~\ref{fig:setup}. A mode-locked laser source (Gaussian mode) with a wavelength of 355 nm and an average power of 350 mW was used to pump a 3-mm-thick type I BBO crystal to produce collinear, degenerate entangled photon pairs via SPDC. Using a  $4f$ telescope, the plane of the crystal was imaged (2$\times$) onto two separate SLMs which are encoded with the BG transmission functions. The SLM planes were re-imaged (0.4$\times$) by a $4f$ telescope and coupled into single-mode fibers, which support only the fundamental Gaussian mode. The fibers were connected to avalanche photodiodes, the outputs of which are connected to a circuit that gives the coincidence count rate. 

\begin{figure}[htbp]
\centering
\includegraphics[scale=0.29]{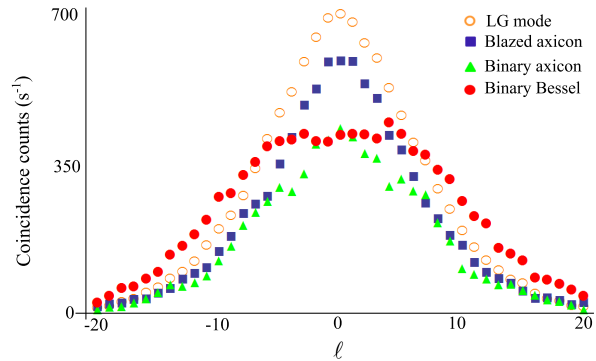}
\caption{Graph of the measured coincidence count rate as a function of OAM for four different measurement schemes. The BG measurements were all measured for $k_{r} = 21$ rad/mm. The empty orange circles represent the measurements recorded for LG modes. The blue squares represent the measurements recorded using a blazed axicon function. The binary axicon function is represented by the green triangles and the measurements from the binary Bessel function are illustrated by red circles.}
\label{fig:KR}
\end{figure}

The OAM bandwidth (also referred to as the spiral bandwidth) was measured as a function of the different BG phase patterns, and compared with the spiral bandwidth measured in the LG basis. Due to conservation of angular momentum \cite{Mair2001}, a coincidence can only be observed for $\ell_{A}+\ell_{B} = 0$, where $\ell_{A}$ and $\ell_{B}$ are the azimuthal indices of the functions encoded in SLM A and B, respectively. These results are shown in Fig.~\ref{fig:KR}. 

The inefficiency of the binary phase pattern results in a decrease in the count rate for both binary phase patterns. Our results show that the binary Bessel phase pattern produces the largest full-width-half-maximum (FWHM) value of 21. The binary axicon function gave a FWHM value of 17, while both the blazed axicon and vortex functions produced OAM spectra with FWHM values of 15. These graphs were all measured for a particular value of $k_{r} = 21$ rad/mm. 
\begin{figure}[htbp]
\centering
\includegraphics[scale=0.27]{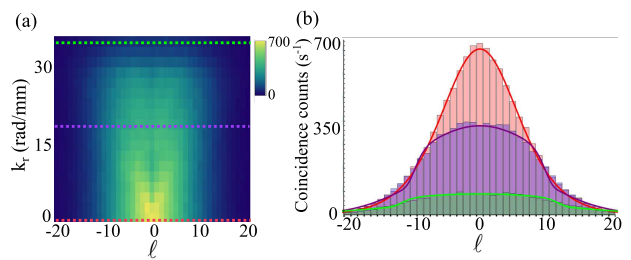}
\caption{(a) Density plot of the modal spectrum in the BG basis for $k_{r}$ and $\ell$. The efficiency of the coincidence count rate decreases as $k_{r}$, however the FWHM of the bandwidth increases with $k_{r}$, seen more clearly in (b). The coloured dashed lines in (a) correspond to the profiles plotted in (b) for $k_{r} = 0$ rad/mm (red), $k_{r} = 21$ rad/mm (purple) and $k_{r} = 35$ rad/mm (green).}
\label{fig:band}
\end{figure}
\\
In investigating the effect of the radial wavevector on the bandwidth, we focused only on the binary Bessel function. We have previously  demonstrated \cite{McLaren2012}, proof of entanglement of such beams, where we have shown a violation of Bell's inequality with the Bell parameter $S = 2.78 \pm 0.05$ for the BG subspace of $\ell = 1$. 
\\
We now illustrate the broadening and flattening of the OAM spectra in the BG basis, as shown in Fig.~\ref{fig:band}. We note that the broadening of the OAM spectrum is at the expense of reducing the coincidence counts at low $\ell$ values. This in turn decreases the heralding efficiency, which has an effect on the security of quantum key distributions. We found, due to the spatial resolution of the SLMs, that there was a maximum limit for which $k_{r}$ could be chosen. We therefore varied the radial wavevector from 0 to 35 rad/mm. We compare the data in Fig.~\ref{fig:band} to the theoretical Schmidt number of Eq.~(\ref{BGcoeff}). 
\\
\begin{figure}[htbp]
\centering
\includegraphics[scale=0.16]{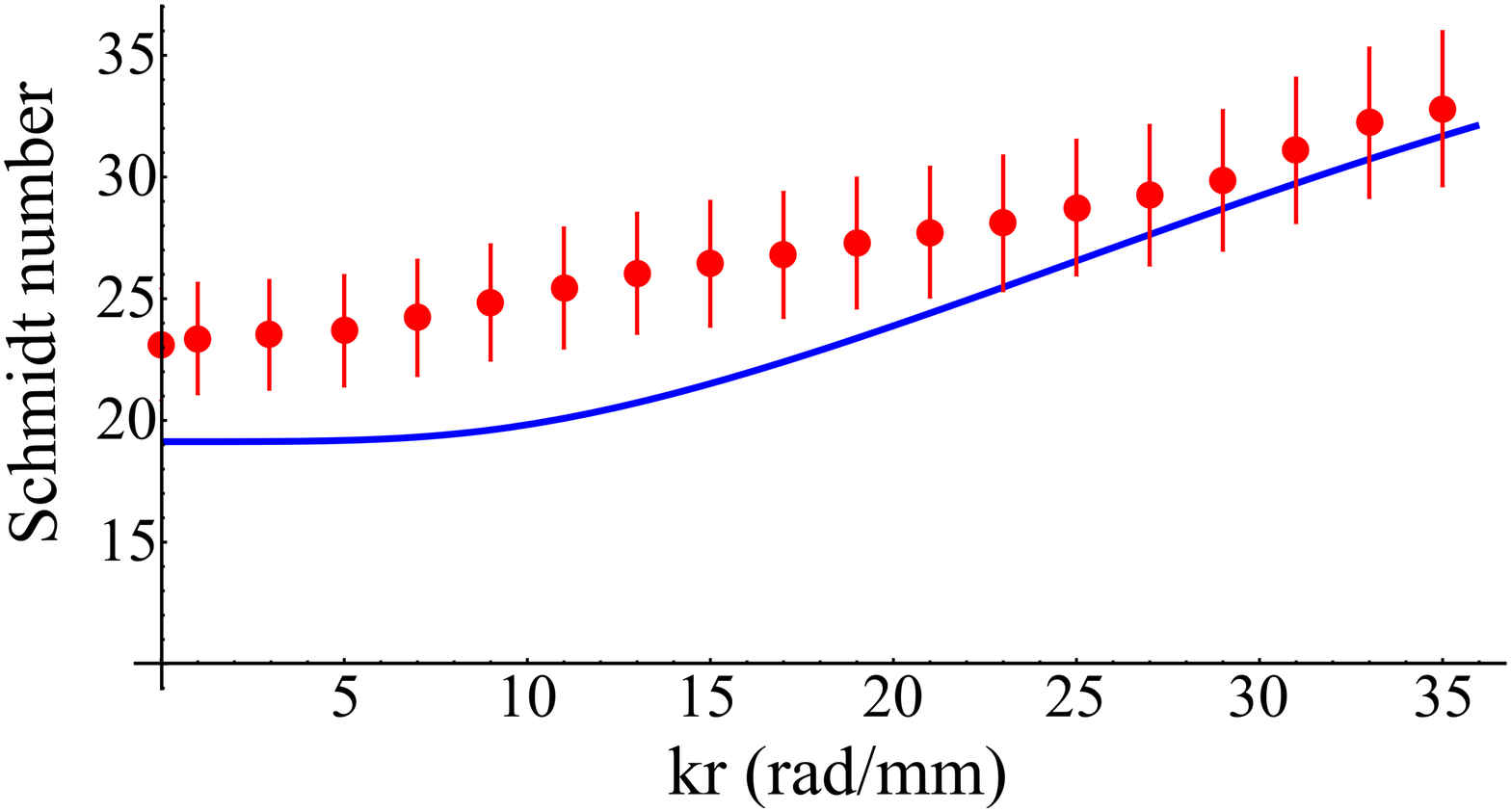}
\caption{Effect of the radial wavevector on the Schmidt number. For $k_{r} = 0$, the transmission function corresponds to that of a vortex mode. An increase in the number of available OAM modes is observed as the radial component is increased. The experimental measurements (red dots) together with a theoretical prediction (solid blue line) are plotted for $\omega_{0} = 0.5$ mm and $\omega_{1} = 0.23$ mm.}
\label{fig:Schmidt}
\end{figure}
\\
In the case of BG modes, the Schmidt number is dependent on the value of the radial wavevector, as seen in Fig.~\ref{fig:Schmidt}. The experimental results are plotted together with a theoretical prediction based on Eq.~(\ref{BGcoeff}) and Eq.~(\ref{schmidt}) for a collinear SPDC setup ($\omega_{0} = 0.50$ mm, $\omega_{1} = 0.23$ mm). 
It is clear that as the value of the radial wavevector increases, so too, does the Schmidt number. These results are reminiscent of entanglement concentration, where maximally entangled states are extracted from non-maximally entangled pure states \cite{Vaziri-2003}. The increase in accessible OAM modes is advantageous for high-dimensional entanglement, hence offering realisable applications in quantum information processing.

\section{Conclusion}
We have used back-projection as an aid in designing a measurement scheme for probing OAM correlations. By propagating a classical beam from one of the detectors onto the plane of the crystal, we were able to examine three transmission functions for generating and measuring modes with both helical and radial structures. We investigated the efficacy of a blazed axicon, a binary axicon and a binary Bessel function in generating BG modes. Only the binary Bessel function, resulted into Bessel-Gauss modes at the plane of the crystal. OAM correlations in photons generated via SPDC were then measured using these three transmission functions. The OAM bandwidth obtained from the use of both the blazed and binary axicon transmission functions were similar to the OAM bandwidth obtained when using spiral phase masks, which measure LG modes. However, using the binary Bessel transmission function to measure BG modes, leads to a larger OAM bandwidth and more usable OAM modes. The number of modes were shown to increase in a tunable manner, with the minimum set by the LG case. This is useful for quantum information and communication applications requiring entanglement in higher dimensions.

\end{document}